\documentclass[prb,twocolumn,aps,showpacs]{revtex4}
\usepackage{graphicx,epsfig}
\usepackage{bm}
\usepackage{amssymb} 
\begin{document}

\title{Quantum dynamics of nuclear spins and spin relaxation in organic semiconductors}

\author{V. V. Mkhitaryan and V. V. Dobrovitski}

\affiliation{Ames Laboratory, Iowa State University, Ames, Iowa
50011, USA}

\begin{abstract}

We investigate the role of the nuclear spin quantum dynamics in
hyperfine-induced spin relaxation of hopping carriers in organic
semiconductors. The fast hopping regime with a small carrier spin
precession during a waiting time between hops is typical for
organic semiconductors possessing long spin coherence times. We
consider this regime and focus on a carrier random walk diffusion
in one dimension, where the effect of the nuclear spin dynamics
is expected to be the strongest. Exact numerical simulations of
spin systems with up to 25 nuclear spins are performed using the
Suzuki-Trotter decomposition of evolution operator. Larger nuclear
spin systems are modeled utilizing the spin-coherent state
$P$-representation approach developed earlier. We find that the
nuclear spin dynamics strongly influences the carrier spin
relaxation at long times. If the random walk is restricted to a
small area, it leads to the quenching of carrier spin polarization
at a non-zero value at long times. If the random walk is
unrestricted, the carrier spin polarization acquires a long-time
tail, decaying as $ 1/\sqrt{t}$. Based on the numerical results,
we devise a simple formula describing the effect
quantitatively.
\end{abstract}


\maketitle

\section{Introduction}

One of the most remarkable characteristics of organic
semiconductors is long spin coherence times, which makes these
materials suitable for various applications in spintronics
\cite{Naber07, Sanvito07, VardST10} and magnetotransport devices.
\cite{Dediu02, VardShi, Wohlgen06, BobHypExp, Dediu09, ThoNatMat,
Molenkamp11, Molenkamp13, Dediu13} The reason behind this valuable
property is that the light elements as hydrogen and carbon, from
which organic semiconductors are composed, have very weak
spin-orbit interaction. For the same reason, the hyperfine
coupling between the charge carrier and hydrogen nuclear spins is
often the dominant source of the carrier spin scattering,
\cite{BobHypExp, Dediu09, ThoNatMat, Shinar, LupBoePRL, Malissa}
and the resulting spin relaxation.

Due to the inherently present disorder, charge transport in
organic semiconductors occurs via incoherent diffusive random walk
over the charge carrying molecules or $\pi$-conjugated segments of
polymers. During the waiting time between two consecutive hops the
carrier spin and the local hydrogen nuclear spins couple by a
hyperfine interaction. Existing theories of hyperfine-induced spin
relaxation and more general spin-dependent phenomena in organic
semiconductors approximate the hyperfine interaction by locally
random static magnetic fields, acting on the carrier spin.
\cite{SchulWol, BobSpinDiff, FlattePRL12, RR13, RaikhSpinRel, we}
This provides a semiclassical approximation to the quantum spin
dynamics. The distribution of random magnetic fields is taken to
be Gaussian. The hyperfine interaction strength is then controlled
by the mean square deviation of the Gaussian distribution,
$b_{\text{hf}}$. Experimentally established values of
$b_{\text{hf}}$ are of order of few mT.

The semiclassical approximation neglects the action of the carrier
spin on the local nuclear spin environment.
Thus it is good for relatively slow nuclear spin dynamics and fast
carrier hopping. However, the semiclassical approximation may be
inadequate if the carrier spends long times at the molecular
sites, leading to sizeable changes in the local spin environments.
This can happen, e.g., if the carrier moves in effectively lower
dimensions, so that its random walk trajectory undergoes multiple
self-intersections. In the case of one-dimensional diffusive
random walk of a total duration $t$, the average time spent by a
carrier on a site on its trajectory is $\sim\sqrt{t\tau_0}$, and
in $d=2$ dimensions this time is $\sim \tau_0 \ln(t/\tau_0)$,
where $\tau_0$ is the average waiting time, i.e., the average time
between consecutive hops. \cite{MontWeiss} In both these cases the
total time spent by the carrier at a site can become sufficiently
long for the quantum dynamics of local spin environment to be
important.

In this paper we investigate the effect of nuclear spin quantum
dynamics on the spin relaxation in $d=1$. We utilize numerical
simulations based on a Monte Carlo sampling of random walk
trajectories. To simulate the spin dynamics of small systems with
up to $25$ total number of nuclear spins we employ the
Suzuki-Trotter decomposition of evolution operator. \cite{ST1,
ST2} For systems with larger number of nuclear spins we make use
of the coherent-state $P$-representation approach for quantum
central spin dynamics. \cite{SlavaRev, SlavaPrep}

We study the time decay of average spin polarization of a carrier
spin, $P(t)$, initially injected into an organic layer in the
spin-up state [$P(0)=1$]. The spin polarization evolves as the
carrier walks randomly over a linear chain of $L$ molecular sites,
and its spin interacts with $N$ nuclear spins at each site. The
decay of $P(t)$ is controlled by the dimensionless combination,
$\eta=(b_{\text{hf}}\tau_0)$, \cite{BobSpinDiff, RaikhSpinRel, we}
which is the average precession angle of carrier spin between two
consecutive hops (we take $\hbar=1$ and the electron gyromagnetic
ratio, $\gamma_e=1$, making the magnetic fields and the Larmor
frequencies equivalent). We focus on small $\eta\ll1$,
characteristic for organic semiconductors featuring long spin
coherence time.

Our calculations prove that the initial, dominant decay of spin
polarization, down to about $P(t)=0.05$, is insensitive to the
dynamics of nuclear spins and can be obtained very accurately from
the semiclassical approximation. Therefore the results of previous
studies based on the semiclassical approach \cite{RaikhSpinRel,
we} remain valid for the initial decay of $P(t)$.

The effect of quantum dynamics of nuclear spins develops at longer
times. For relatively small $L$ and $N$, we observe a plateau-like
long-time behavior of $P(t)$, where its value is about
$(LN)^{-1}$, independently from $\eta$. This is in sharp contrast
to the semiclassical behavior. For finite $N$ and $L\to\infty$ we
find yet another unique long-time dependence, namely
$P(t)\approx\alpha\bigl(N\sqrt{t/\tau_0}\bigr)^{-1}$, where
$\alpha\approx0.43$. On the other hand, for finite $L$ and
increasing $N$, the long-time polarization disappears and $P(t)$
simply regains the semiclassical form.

The paper is organized as follows. In the next Section we discuss
the hyperfine interaction between the carrier and hydrogen nuclear
spins, and its semiclassical description in terms of random static
magnetic fields. Our results on the spin relaxation by a quantum
nuclear spin bath and their comparison with those of the
semiclassical approximation are presented in
Sect.~\ref{secCalcul}. We discuss our results and provide
explanations for the common features and differences of the
quantum and semiclassical results in Sec.~\ref{secDiscuss}.


\section{The hyperfine interaction and its semiclassical description}

We consider a diffusive random walk of a carrier over a linear
chain of $L$ molecular sites. Accordingly, the sites are
enumerated by the scalar coordinate, $r=1,..,L$. At each molecular
site $r$, the carrier spin $\mathbf{S}=1/2$ couples to
$k=1,2,..,N$ nuclear spins $\mathbf{I}_{ r, k}=1/2$ by a hyperfine
interaction. We will consider an isotropic hyperfine coupling,
governed by the Hamiltonian
\begin{equation}\label{locHam}
H_{r}=B_0 S_z + {\bf S}\,\sum_{k=1}^N a_k {\bf I}_{r k},
\end{equation}
where $B_0$ is the external magnetic field along the $z$-axis, and
$\{a_k\}_{k=1}^N$ are the hyperfine coupling constants. This
description implies that the carrier-host molecular sties are
identical, so that $a_k$ do not depend on $r$.

Quite generally, theories of spin related phenomena in organic
semiconductors approximate the quantum Overhauser field, given by
the sum in Eq. (\ref{locHam}), by a constant classical vector,
\begin{equation}\label{Bi}
{\bf b}_r=\sum_{k=1}^Na_k \mathbf{I}_{r k},
\end{equation}
which is not affected by the interaction.  \cite{SchulWol} The
approximation Eq. (\ref{Bi}) can be justified as follows. With the
interaction Eq. (\ref{locHam}), the nuclear spin precession period
scales as $\sqrt{N}$. \cite{MerkEfRos} Therefore, for large $N$
one can neglect the slow dynamics and consider the nuclear spins
being static. For large $N$ it is also reasonable to approximate
the distribution of ${\bf b}_r$ by the Gaussian with the standard
deviation, $b_{\text{hf}}= \frac12 \sqrt{\sum_ka_k^2}$.

While one expects that the approximation Eq. (\ref{Bi}) is good
for increasingly large $N$, its accuracy for finite $N$ is
questionable. In organic materials $N\approx 10$ is the expected
number of hydrogen nuclei coupled to a carrier spin at a molecular
site. \cite{ThoNatMat, BobSpinDiff} With this large $N$,
the relaxation of a diffusing carrier spin can still be sensitive
to the quantum dynamics of nuclear spins.

\section{Calculation of spin relaxation caused by a quantum nuclear spin bath}


\label{secCalcul}

In this Section we calculate the spin relaxation of a carrier
performing a diffusive random walk over a linear chain of $L$
sites. At each site the carrier spin couples to $N$ nuclear spins
according to the Hamiltonian (\ref{locHam}). We assume that the
random hops occur between the nearest neighbor sites with equal
probability. Accordingly, the random waiting times have the
average, $\tau_0$, uniform for all sites and the waiting time
distribution at each site is Poissonian,
$\mathcal{N}(\tau)=\tau_0^{-1}\exp(-\tau/\tau_0)$. For finite $L$,
boundary conditions for the random walk should also be specified.
However, simulations with periodic or reflective boundaries yield
very close results, so that we present the ones with periodic
boundary conditions.

Our calculation of the spin relaxation is based on the following
simple consideration. A carrier, initially injected at $r=r_0$ in
the spin-up state $|\uparrow\rangle$, diffusively moves over the
available molecular sites and suffers spin flips. The system thus
begins its evolution from the initial spin state,
$|\psi(0)\rangle= |\uparrow\rangle\otimes|\chi \rangle$, where
$|\chi\rangle$ is the initial wavefunction of all $(L\cdot N)$
nuclear spins. In the course of carrier random walk with the
trajectory $r(t)$,
the spin state of the system evolves according to the
Schr\"{o}dinger equation, $i\partial_t|\psi(t )\rangle =
H_{\mathbf{r}(t)} |\psi(t)\rangle$. Its solution can be formally
written in terms of the time-ordered exponent,
\begin{equation}\label{stvec}
|\psi(t, [r(t)])\rangle= T\exp\bigl( -i\int_0^tdt'
H_{\mathbf{r}(t')}\bigr) |\uparrow\rangle\otimes|\chi \rangle.
\end{equation}
By writing $[r(t)]$ in the argument we emphasize that $|\psi(t,
[r(t)])\rangle$ is the spin wavevector after time $t$, provided
that the carrier underwent a random walk with the trajectory
$r(t)$.
The carrier spin polarization can be defined as $P(t)=
2\overline{S_z(t)}$ with the bar standing for a triple average,
where the first is the quantum mechanical average, the second is
the average over the possible quantum states of the nuclear spin
bath, and the third extends over different realizations of random
walk trajectories. Thus
\begin{equation}\label{trpav}
P(t)= 2\Bigl \langle \Bigl \langle \langle \psi(t, [r(t)]
)|S_z|\psi(t, [r(t)]) \rangle \Bigr \rangle_{\!\!\{\chi\}}
\Bigr\rangle _{\text{rw}}.
\end{equation}

The calculation of $P(t)$ from Eq. (\ref{trpav}) can be carried
out numerically by utilizing the Suzuki-Trotter decomposition
\cite{ST1, ST2, SlavaRev} of the time-ordered exponent in
Eq.~(\ref{stvec}), combined with a Monte Carlo sampling of random
walk trajectories. However, as is well known, memory requirements
impose severe limitations on the number of spins that can be
modeled. \cite{SlavaRev} This is because the required memory grows
exponentially with the number of spins. In our case this means
that the total number of nuclear spins, $(L\cdot N)$, is
restricted to only few tens.

\begin{figure}[t]
\vspace{-0.3cm}
\centerline{\includegraphics[width=95mm,angle=0,clip]{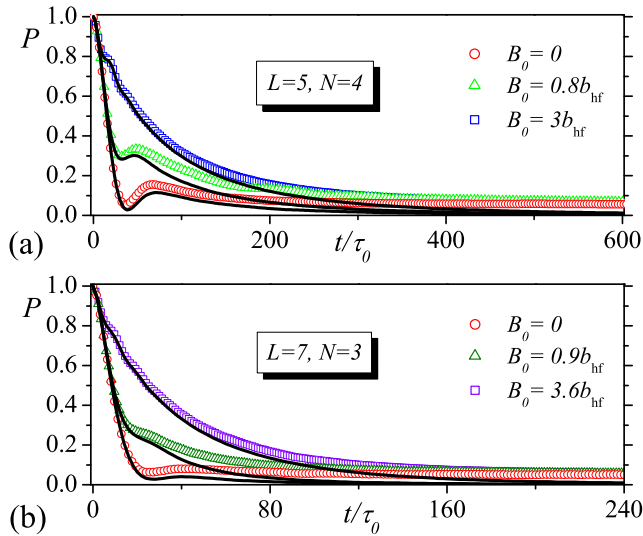}}
\vspace{-0.4cm} \caption{(Color online) Spin relaxation in various
applied magnetic fields, obtained by the numerically exact
simulation of quantum spin dynamics. $P(t)$ is plotted with open
symbols. Black lines are the results from the semiclassical
approximation. (a): System of $L=5$ sites, $N=4$ nuclear spins per
site, with $\eta \equiv(b_{\text{hf}}\tau_0) =0.1$ and the
hyperfine couplings, $a_1=0.83b_{\text{hf}}$,
$a_2=0.9b_{\text{hf}}$, $a_3=1.05b_{\text{hf}}$,
$a_4=1.18b_{\text{hf}}$. Plotted are the results for relaxation in
the applied fields, $B_0=0$ (red), $B_0=0.8b_{\text{hf}}$ (green),
and $B_0=3b_{\text{hf}}$ (blue). (b): System of $L=7$ sites and
$N=3$ nuclear spins per site, with $\eta=0.167$ and
$a_1=1.3b_{\text{hf}}$, $a_2=1.15 b_{\text{hf}}$,
$a_3=0.99b_{\text{hf}}$. Red, dark green, and violet symbols
represent the applied fields, $B_0=0$, $B_0=0.9b_{\text{hf}}$, and
$B_0=3.6b_{\text{hf}}$, respectively.
} \label{quantsemi}
\end{figure}

\begin{figure}[b]
\vspace{-0.3cm}
\centerline{\includegraphics[width=95mm,angle=0,clip]{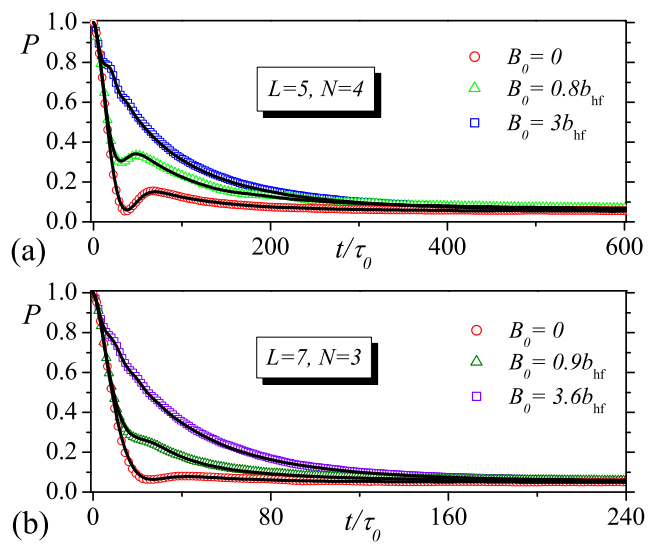}}
\vspace{-0.4cm} \caption{(Color online) Numerically exact
calculation results from Fig.~\ref{quantsemi} (open symbols) are
plotted together with the results obtained from the
$P$-representation method (black lines). The comparison confirms
the high accuracy of the $P$-representation approach to this
problem.} \label{comparison}
\end{figure}

Based on the above scheme, we have simulated the quantum spin
dynamics of various systems with up to $25$ total number of
nuclear spins, uniformly distributed at $L=3$--25 molecular sites.
Representative results of such simulations, for a system of $L=5$
sites and $N=4$ spins per site with
$\eta\equiv(b_{\text{hf}}\tau_0)=0.1$, and of $L=7$ sites and
$N=3$ spins per site with $\eta=0.167$, are demonstrated in
Figs.~\ref{quantsemi} (a) and (b), respectively. The plots also
contain numerical results for the same systems, found from the
semiclassical approximation Eq. (\ref{Bi}) [for the details of
semiclassical calculation of spin relaxation see Ref.
\onlinecite{we}]. At small $t$, the quantum and semiclassical
results in Fig.~\ref{quantsemi} coincide. This indicates that the
initial decay of $P(t)$ is nearly insensitive to the quantum
dynamics of nuclear spins, so that it can be well described
semiclassically. On the other hand, there is a strong discrepancy
between the quantum and semiclassical results at longer times.
Thus the true long-time behavior of $P(t)$ is not captured by the
semiclassical approximation; while the latter predicts a vanishing
spin polarization, the decay of $P(t)$ shows a long-time plateau
behavior, quenching at a small but finite value.

Definitely, the quenching of $P(t)$ at long times, seen in
Fig.~\ref{quantsemi}, is a consequence of the nuclear spin
dynamics. However, further analysis of the effect of nuclear spin
dynamics requires a modeling of larger spin systems. To this end
we employ the $P$-representation method for quantum spin system
simulation, proposed in Ref. \onlinecite{SlavaRev} (see also Ref.
\onlinecite{SlavaPrep}). The method is based on a Monte Carlo
sampling of the density matrix in the spin coherent-sate basis,
allowing an efficient modeling of large quantum systems with
thousands of bath spins.
It also allows to assess the spin dynamics at considerably longer
times.

To check the accuracy of the $P$-representation method for the
problem at hand, we compare its outcome for systems of up to $25$
nuclear spins with the results of exact simulation. The comparison
clearly verifies the efficiency of $P$-representation approach to
our problem. We demonstrate typical examples of such comparison in
Fig.~\ref{comparison}, by plotting the quantum simulation results
of Fig.~\ref{quantsemi} together with the $P$-representation
curves for the same systems.
The $P$-representation curves in Fig.~\ref{comparison} virtually
coincide with the exact simulation results both in short- and
long-time domains, and only a minor deviation can be observed at
the intermediate times.

\begin{figure}[t]
\vspace{-0.3cm}
\centerline{\includegraphics[width=95mm,angle=0,clip]{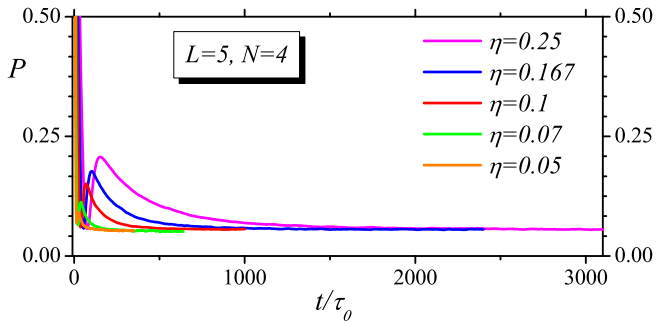}}
\vspace{-0.4cm} \caption{(Color online) Long-time behavior of
$P(t)$ for $L=5$, $N=4$, and
$\eta=0.25$ (orange), $\eta=0.167$ (green), $\eta=0.1$ (red),
$\eta=0.07$ (blue), and $\eta=0.05$ (magenta). } \label{diffeta}
\end{figure}

With the $P$-representation method at hand we are able to
investigate both larger spin systems and longer time dynamics. In
particular, we examine the long-time behavior of the spin
relaxation in the regime of small $\eta\ll1$, where the relaxation
is slow. Our $P$-representation simulations reinforce the previous
conclusion that at long times the decay of $P(t)$ is very slow, so
that the spin polarization nearly quenches to a few percents of
its original value. We also find that the value of $P(t)$ at the
long-time plateau is almost insensitive to $\eta$. This is shown
in Fig.~\ref{diffeta}, where $P(t)$ is plotted for various values
of $\eta$, for the system with $L=5$, $N=4$. Moreover, as seen in
Figs.~\ref{quantsemi} and \ref{comparison}, the plateau value of
$P(t)$ is almost independent on the external magnetic fields of
the order of $b_{\text{hf}}$. Therefore the plateau value is
universal in the sense that it is determined by $L$ and $N$, and
does not depend on $\eta$ and $B_0$.
\begin{figure}[b]
\vspace{-0.3cm}
\centerline{\includegraphics[width=95mm,angle=0,clip]{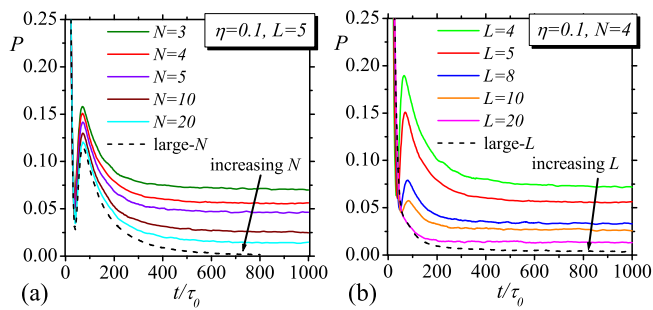}}
\vspace{-0.4cm} \caption{(Color online) Dependence of the
long-time plateau value of $P(t)$ on the number of molecular
sites, $L$, and number of spins per site, $N$, for $\eta=0.1$. (a)
$P(t)$ is plotted for $L=5$ and five different $N$ ranging from
$3$ to $20$. For $N\to \infty$, the curves saturate at the dashed
line, which is the semiclassical approximation result for the same
$L$. (b) $P(t)$ is plotted for five different $L$-values ranging
from $4$ to $20$, and $N=4$ fixed. The dashed line is the
saturation curve, $N=4$, $L\to \infty$.} \label{diffLandN}
\end{figure}

Further numerical analysis of the long-time dynamics shows that
the amplitude of the long-time plateau of $P(t)$ decreases both
with $L$ and $N$. Specifically, for not too large $L$ and $N$ the
amplitude of the long-time plateau scales as $(LN)^{-1}$
(excluding $L=1$). For $\eta=0.1$, we demonstrate this dependence
in  Fig.~\ref{diffLandN}. For fixed $L$ and increasing $N$, $P(t)$
eventually saturates at the curve obtained from the semiclassical
description Eq.~(\ref{Bi}) [see Fig.~\ref{diffLandN}(a)]. This
saturation means a total elimination of the effect of nuclear spin
dynamics for large $N$. The decay regime of $P(t)$ with a fixed
small $N$ and increasingly large $L$ is less obvious. The
saturation curve of $P(t)$ in Fig.~\ref{diffLandN}(b) retains the
long-time tail and thus differs from the result of semiclassical
approximation qualitatively. This is an indication that the
long-time behavior of $P(t)$ remains sensitive to the quantum
dynamics of nuclear spins for $L\to\infty$.

\begin{figure}[t]
\vspace{-0.3cm}
\centerline{\includegraphics[width=95mm,angle=0,clip]{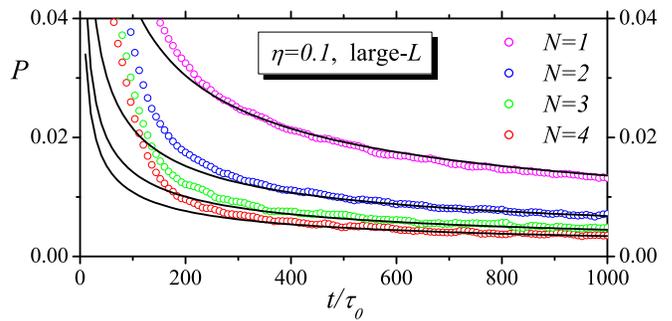}}
\vspace{-0.4cm} \caption{(Color online) Large-L saturation curves
for $\eta=0.1$ and  $N=1$ (magenta), $N=2$ (blue), $N=3$ (green),
$N=4$ (red). Black lines are plotted from Eq.~(\ref{tailVsLN}).
The statistical error bars are of the order of the symbols size. }
\label{Lsatur}
\end{figure}

To understand this effect, we investigate the large-$L$ saturation
curves for different $N$.
We infer that in this regime the long-time tail of $P(t)$ slowly
decays with time as $1/\sqrt{t}$, and scales with $N$ as $1/N$.
Figure \ref{Lsatur} illustrates this dependence for $\eta=0.1$. We
further checked that this result is almost insensitive to $\eta$,
at least for $\eta=0.2$ -- $0.01$. However, we note that the
overall amplitude of the effect is small; for $N=2$, for example,
the $1/\sqrt{t}$ decay sets up at $P(t)\approx 0.012$ (see
Fig.~\ref{Lsatur}).

\section{Discussion}

\label{secDiscuss}

As we have shown above, the quantum dynamics of nuclear spins does
not influence the initial relaxation of $P(t)$ but shows up at
long times, as a plateau for finite $L\cdot N$, or as a slow decay
$\sim 1/\sqrt{t}$ for $L\to\infty$.

The fact that the initial decay of $P(t)$ is not affected by the
nuclear spin dynamics can be understood from the following
reasoning. Defining the cumulant expansion,
$P(t)=\exp\bigl[\sum_{n=1}^\infty K_n(t)\bigr]$, where $K_n\propto
b_{\text{hf}}^n$, from Eq. (\ref{trpav}) one can find the first
non-vanishing cumulant, $K_2$, to be
\begin{equation}\label{K2}
K_2(t)=-2b_{\text{hf}}^2\int\limits_0^t\!dt_1\!
\int\limits_0^{t_1}\!dt_2 G_L (0, t_1-t_2),
\end{equation}
where $G_L(r, t)$ is the Green function of the random walk over
$L$ sites. Equation (\ref{K2}) exactly coincides with the second
cumulant function which can be found from the semiclassical
approximation. \cite{we} This explains why the initial relaxation
is insensitive to the quantum dynamics of nuclear spin, even in
the ultra-quantum case $N=1$.

Similarly to the semiclassical result, \cite{we} the approximation
$P(t)\approx\exp\bigl[K_2(t)\bigr]$ correctly describes the
dominant spin relaxation, so that the spin relaxation time,
$\tau_S$, is set by $K_2(t)$ and is insensitive to the nuclear
spin dynamics. For large number of sites ($L\gg\eta^{-2/3}$) the
form $P(t)\approx\exp\bigl[-(t/\tau_S)^{3/2}\bigr]$ with
$\tau_S\simeq \tau_0/\eta^{4/3}$ follows from the results of previous
semiclassical treatments. \cite{RaikhSpinRel, we}

Our findings for the long-time behavior of $P(t)$ can be
summarized in the formula,
\begin{equation}\label{tailVsLN}
P(t)\approx\frac1N\left(\frac1L+
\frac\alpha{\sqrt{t/\tau_0}}\right),\qquad  t\gg\tau_S,
\end{equation}
where $\alpha\approx 0.43$ is a constant. The combination in the
brackets of Eq.~(\ref{tailVsLN}) is approximately the inverse
number of sites visited by the carrier. Indeed, for small $L$ and
$t/\tau_0\gg L^2$, all the $L$ sites are visited equally many
times and the inverse number of visited sites is $1/L$. For
$L\to\infty$, on the other hand, the average number of sites
visited by the carrier is $\sim\sqrt{t/\tau_0}$, and thus its
inverse, $\alpha/\sqrt{t/\tau_0}$. Then Eq.~(\ref{tailVsLN}) is
the inverse number of nuclear spins that couple to
the carrier spin during a long-term diffusion.

To further elucidate this behavior we note that the Hamiltonian
(\ref{locHam}) conserves the $z$-component of the total spin.
Therefore the initial carrier spin polarization does not average
to zero but gets redistributed between the carrier and nuclear
spins. It is reasonable to expect that for systems with smaller
number of nuclear spins $P(t)$ can be essentially non-zero for
arbitrarily long times. However, the fact that this redistribution
leads to a non-oscillatory, fixed or slowly changing $P(t)$ at
long times is highly nontrivial. Moreover, the peculiar dependence
Eq.~(\ref{tailVsLN}) means that the polarization becomes evenly
distributed between the carrier spin and nuclear spins which
couple to the carrier spin in the course of diffusion.

It is also remarkable that the amplitude of the tail is almost
independent of $\eta$, but is determined by $L$ and $N$. This
independence resembles the central spin problem ($L=1$, large
$N$), where the central spin polarization evolves into $1/3$ of
its initial value regardless of the interaction strength,
\cite{SchulWol, MerkEfRos} and relaxes from this value very
slowly. \cite{ErlNaz, SlavaPrep, SlavaRev}

In conclusion, we have demonstrated that the quantum dynamics of
nuclear spins leads to the long-time steady or slowly decreasing
carrier spin polarization; a feature which is not captured by the
semiclassical approximation. The long-time behavior extends up to
the times when other relaxation mechanisms become important (e.g.,
carrier spin-lattice relaxation times, or hydrogen nuclear spin
dephasing times). The effect can be
strong for carriers diffusing over fewer molecular sites, e.g., in
the situation realizing in tunnel magnetoresistance experiments in
organic spin valves, at the onset of multiple-step tunneling.
\cite{BobHypExp}

\section*{Acknowledgments}

We thank J. Shinar and M. E. Raikh for helpful discussions. Also
we are grateful to H. Terletska for reading the manuscript. Work
at the Ames Laboratory was supported by the US Department of
Energy, Office of Science, Basic Energy Sciences, Division of
Materials Sciences and Engineering. The Ames Laboratory is
operated for the US Department of Energy by Iowa State University
under Contract No. DE-AC02-07CH11358.





\end{document}